# Toward a reliable PWM-based light-emitting diode visual stimulus for improved SSVEP response with minimal visual fatigue

*Surej Mouli, Ramaswamy Palaniappan*

**Abstract:** Steady-state visual evoked response (SSVEP) is widely used in visual-based diagnosis and applications such as brain–computer interfacing due to its high information transfer rate and the capability to activate commands through simple gaze control. However, one major impediment in using flashing visual stimulus to obtain SSVEP is eye fatigue that prevents continued long-term use preventing practical deployment. This combined with the difficulty in establishing precise pulse-width modulation (PWM) that results in poorer accuracy warrants the development of appropriate approach to solve these issues. Various studies have suggested the usage of high frequencies of visual stimulus to reduce the visual fatigue for the user but this results in poor response performance. Here, the authors study the use of extremely high duty-cycles in the stimulus in the hope of solving these constraints. Electroencephalogram data was recorded with PWM duty-cycles of 50–95% generated by a precise custom-made light-emitting diode hardware and tested ten subjects responded that increasing duty-cycles had less visual strain for all the frequency values and the SSVEP exhibited a subject-inde-pendent peak response for duty-cycle of 85%. This could pave the way for increased usage of SSVEP for practical applications.

## 1 Introduction

Steady-state visual evoked potential (SSVEP) is elicited in the brain when a user attentively gazes at a visual stimulus flashing at a constant frequency. SSVEP is a periodic response generated in the occipital region of the brain when the user focuses on a flashing visual stimulus and would have the same frequency as the presented visual stimulus [1]. SSVEP is widely used in academic and clinical research studies due to the minimal amount of training required and also as it allows a higher signal-to-noise ratio [2–4]. SSVEP-based electroencephalogram (EEG) responses are used in medical diagnosis for investigating visual perceptions, user attention or brain responses to identify any symptoms or other medical conditions [5]. Researchers have also explored SSVEP application in emotion and electrophysiological study to explore the face inversion effect and the N170 potential [6, 7]. Studies have also explored the application of SSVEP responses for infants in investigating visual acuity and cortical functions [8, 9]. SSVEP uses non-muscular communication channel and that makes it widely acceptable in brain–computer interface (BCI) applications, where it can support people with disabilities to control external application with multiple visual stimuli [10–13]. Other form of BCIs includes hybrid versions, which combines paradigms such as P300, visual or audio to improve the accuracy and efficiency in BCI operations [14–16]. Fig. 1 shows an EEG data acquisition and processing block diagram for an SSVEP-based system.

SSVEP is a repetitive response generated in the brain and is synchronised with the frequency of the visual stimulus. To produce an accurate response, the stimulus flicker frequency must be precise and consistent throughout the period the user is focusing. Traditionally, SSVEP is evoked using flickers produced with liquid crystal display (LCD) screens in which the flicker frequencies are limited to the refresh rate of the LCD [17–19]. For a standard LCD screen, the refresh rate is fixed at 60 Hz and the frequencies that can be generated are 6.66 Hz (i.e. 9 fps, 60/9), 7.5, 8.57, 10, 12 and 15 Hz. It is not possible to generate any other required frequency such as 7, 8 or 9 Hz or choose high duty-cycles of the visual flicker due to the fixed refresh rate of the screen. Gazing the visual stimulus on an LCD screen for longer periods of time can make the user tired due to visual fatigue not only from the flashing objects on screen but also from the flicker generated by the screen itself which can reduce the SSVEP response. Even though the eye strain can be reduced by using higher-frequency flickers, it reduces the SSVEP amplitudes resulting in weaker responses. On the other hand, SSVEP responses are higher for lower-frequency ranges but would introduce visual fatigue for prolonged usage [20, 21].

Studies have used light-emitting diode (LED)-based visual stimulus designs that can produce visual flickers without any frequency restrictions and can also produce multiple visual stimuli using the same control platform [22, 23]. When using multiple visual stimuli flashing at different frequencies, the accuracy of the flicker frequency for visual stimulus plays an important role in SSVEP-based BCI. Even though the frequency restriction is resolved with LED usage, the visual fatigue and user comfortability are still cause for concern. To mitigate these issues, this paper uses a visual stimulus hardware based on LEDs which is capable of producing any frequency visual flicker with adjustable duty-cycle using pulse-width modulation. Duty-cycle can be defined as

$$\text{Duty-cycle} = (T_{\text{ON}}/(T_{\text{ON}} + T_{\text{OFF}})) \times 100\% \quad (1)$$

where $T_{\text{ON}}$ is the stimulus on period and $T_{\text{OFF}}$ is the period when there is no stimulus.

Fig. 2 shows a square-wave with ON and OFF periods marked within a complete cycle. In Fig. 3, examples of square-wave based on various duty-cycles that are used in this paper are shown.

LED-based visual stimulus used in almost all SSVEP studies have used a duty-cycle of 50% with an equal ON/OFF period for the flicker. This when gazed for longer periods of time would cause eye fatigue and lowers the attention and reduces the SSVEP response. In this paper, different duty-cycles of 50, 80, 85, 90 and 95% are analysed in conjunction with lower frequencies of 7, 8, 9 and 10 Hz for reduced visual fatigue and user comfortability.

## 2 Methods

In this research, EEG acquisition system based on Emotiv EPOC+, which is a wireless headset with 14 channels, was used to record the EEG for SSVEP analysis. The EEG data was transmitted wirelessly to the computer by using proprietary communication protocol which monitored the packet loss and electrode contact quality in real time. The raw data acquired was stored in the computer and processed to analyse the SSVEP frequency responses for various duty-cycles. The visual stimulus design was based on chip-on-board (COB) LED in radial form with a diameter of 130 mm emitting green light. COB LEDs form a much denser array producing uniform light and reduce attention shifts. The precise flicker frequencies generated were 7, 8, 9 and 10 Hz using microcontroller with accuracy of ±0.1 Hz which were confirmed at the LED terminals using an oscilloscope. The microcontroller used was based on Teensy development platform which is a 32 bit ARM-based controller and can generate precise square-waves with higher degree of accuracy. This standalone visual stimulus platform could generate any desired frequency with the required duty-cycle which is not possible with traditional LCD screens. For each chosen frequency, the microcontroller was programmed for different duty-cycles 50, 80, 85, 90 and 95% with an accuracy of ±0.1% confirmed with the oscilloscope. Fig. 4 shows a snapshot of the generated square-wave information with 85% duty-cycle.

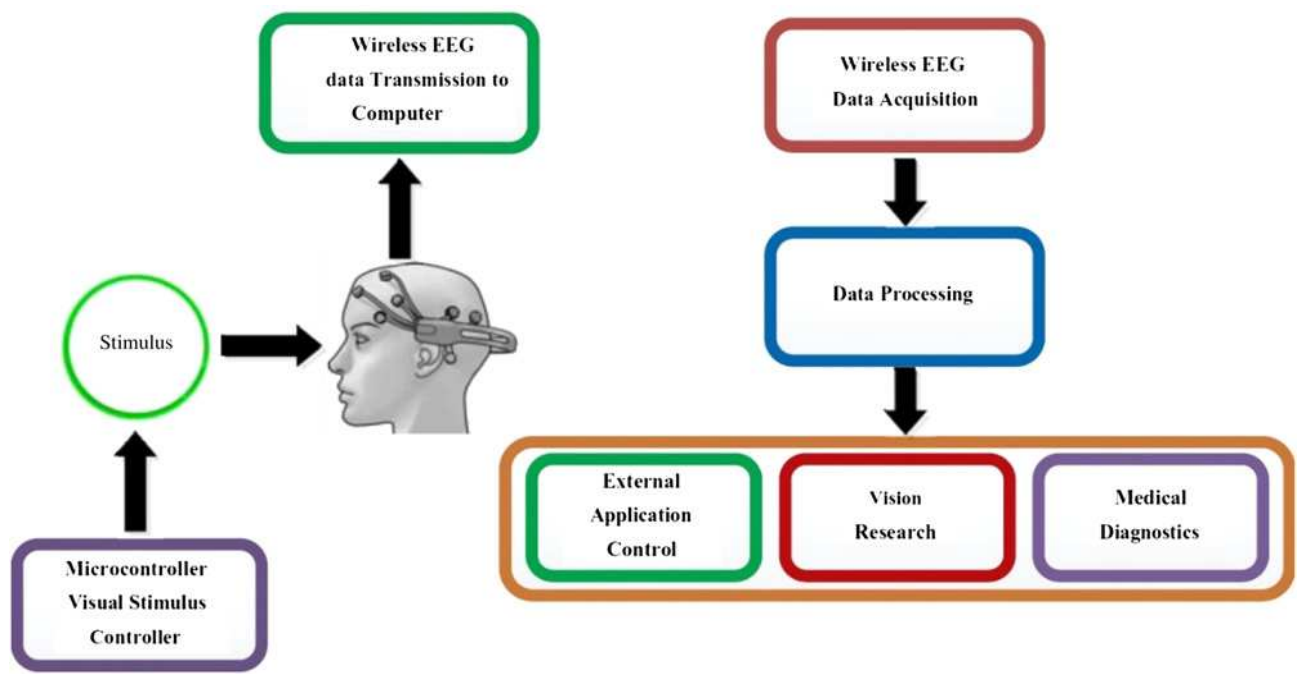

**Fig. 1** *SSVEP-based EEG data acquiring and processing system*

The radial LED was controlled by the microcontroller using a metal–oxide–semiconductor field-effect transistor (MOSFET) driver for constant brightness throughout the experiment. The complete system was powered by a 5 V direct current (DC) source from a battery pack to avoid any mains power line interference while recording the EEG. The EEG responses from all the chosen frequencies and five different duty-cycles were compared and analysed to identify the duty-cycle value that gave the highest response. The experiment also obtained the ratings for the comfort level of the visual stimulus for all the frequency ranges with different duty-cycles.

To explore the influence of duty-cycle in visual flickers, ten participants with perfect or corrected vision within an age group of 25–46 (four females and six males) were chosen for this paper. Participants did not have any prior experience with BCI or any other visual stimulus-based studies. The participants were comfortably seated 60 cm from the visual stimulus which was placed at eye level. Written consents from the participants were obtained after briefing them on the objectives of this paper. Ethical approval was received from Faculty of Sciences Ethics Committee at University of Kent.

The EEG was recorded using Emotiv EPOC+ research edition with 14 electrodes, though only data from a single electrode located at O2 was used in this paper (as SSVEP is maximal in occipital cortex). The visual stimulus was programmed with the desired frequency and duty-cycle to evoke the SSVEP for a period of 30 s for each trial. Each frequency and duty-cycle had five such trials of 30 s for each participant. The EEG headset was prepared with saline soaked felts on all 14 electrodes and positioned on participant's head. The contact quality was verified and the position of the electrodes adjusted to ensure all good contacts using the Emotiv test bench software. EEG recording started with 7 Hz with a randomly chosen duty-cycle for a period of 30 s for five trials followed by other remaining duty-cycles chosen randomly with five trials for the same frequency with 30 s duration. The trials were repeated for 8, 9 and 10 Hz for the same duration and for the five different duty-cycles. Between each 30 s recording session, the participants were given a short break of 1 min to allow any previous stimuli influences to subside. For each participant, the total number of trials was 100 for four frequencies and five duty-cycle values repeated five times.

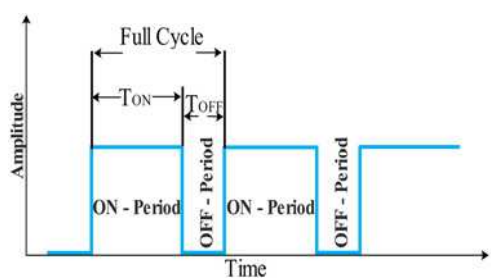

**Fig. 2** *Duty-cycle with ON and OFF periods*

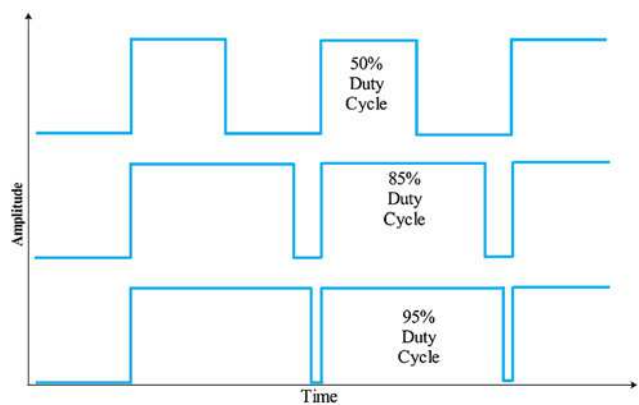

**Fig. 3** *Duty-cycle waveforms with various duty-cycles*

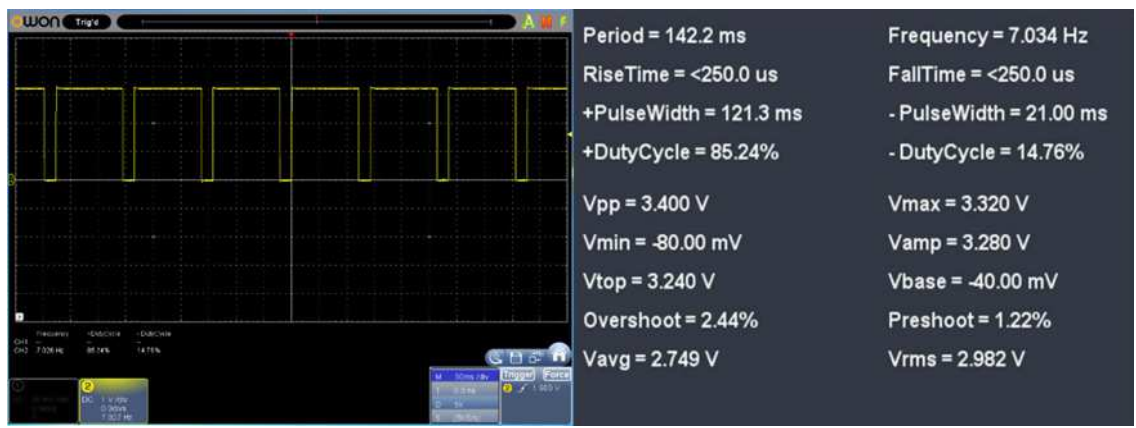

**Fig. 4** *Information snapshot of duty-cycle at 85% for 7 Hz*

### 2.1 Hardware design to generate visual stimulus

The visual stimulus for this paper was designed using a green COB LED ring with a diameter of 130 mm. The colour and the radial size were chosen based on previous studies which identified the highest performance in terms of SSVEP response [20, 24]. The LED ring consists of 156 individual COB LED's densely packed producing a uniform circular green light. The ring was controlled by a 32 bit microcontroller based on Teensy 3.2 development platform. The microcontroller was based on ARM Cortex M4 family which operates at 72 MHz and can generate precise frequencies and duty-cycles. The code for the microcontroller was developed using Arduino software development platform which is easily customisable for any frequency or duty-cycle. The desired visual flicker and duty-cycle were programmed in the microcontroller and the LED was driven through a MOSFET driver to supply the required current for ensuring the constant brightness throughout the experiment. The microcontroller code has to be individually loaded in the microcontroller for each experiment for all the required frequencies and duty-cycles. The hardware platform was powered by a 5 V DC battery pack with 5000 mA current capacity to ensure the steady supply of power to the visual stimulus. Fig. 5 shows the schematic block diagram for the hardware design.

### 2.2 EEG data acquisition and processing

The EEG data was recorded using Emotiv EPOC+ research edition headset with 14 channels and two reference electrodes with the layout as shown in Fig. 6. In this paper, to identify the best duty-cycle for the visual stimulus, only data channel from channel O2 was used for all the different frequency and duty-cycle combinations. Data channel O2 was used since it is in the occipital region and has the maximal response for SSVEP according to various studies [25, 26]. The EEG was recorded in European Data Format (EDF) using Emotiv test bench software for 30 s trials with a sampling frequency of 128 Hz which was fixed in the hardware but sufficient to avoid aliasing as our highest EEG stimulus frequency was only 10 Hz. Each participant had a total recording time of about 50 min (100 trials lasting for 30 s) excluding the rest times of 1 min each between the trials. The recorded data in EDF format for channel O2 was converted to MATLAB format for offline analysis using EEGLAB [27].

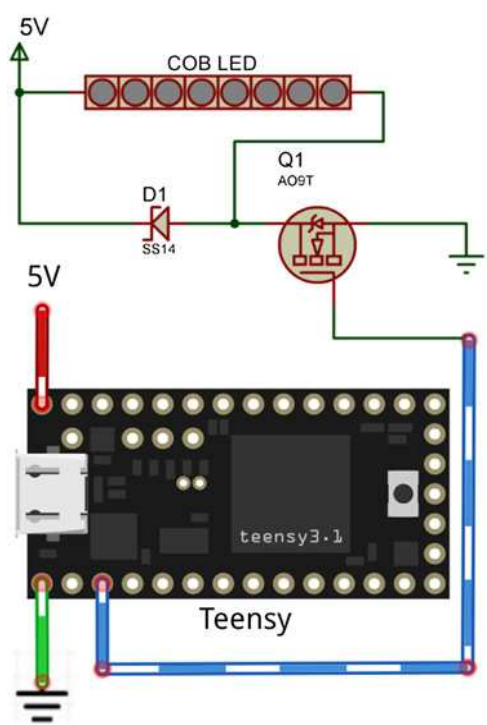

**Fig. 5** *Teensy 3.2 with MOSFET driving radial stimulus*

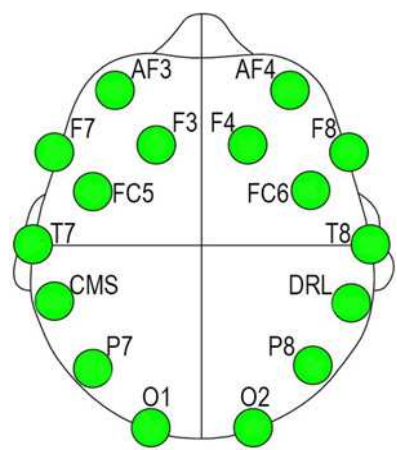

**Fig. 6** *Emotiv electrode location*

The 30 s of EEG data was saved as individual files and filtered using band-pass filter of 2 Hz bandwidth with centre frequency as the stimulus frequency and segmented into 1 s EEG segments. The five trials recorded for each session had 150 segments of 1 s SSVEP EEG data which was analysed using fast Fourier transform (FFT) and the maximum FFT amplitudes stored from each segment for further statistical analysis. Kruskal–Wallis (Normality of data was not established, hence the use of rank-based approaches.) test was used to identify significance of the difference influences of duty-cycles in SSVEP. The analysis was performed on maximal FFT amplitudes of EEG data for all five duty-cycles to identify the most responsive.

## 3 Results

This paper investigated the visual stimulus performance with five different duty-cycles 50, 80, 85, 90 and 95% to compare the SSVEP responses. All the mentioned duty-cycles were analysed for four frequencies 7, 8, 9 and 10 Hz using the COB LED radial visual stimulus. The analysis compared the data from ten participants for identifying the most comfortable and responsive duty-cycle value in all the frequency ranges. For each data set, there were 150 amplitude FFT values for each frequency and duty-cycle from one participant. Each participant had 20 sets of data which included four frequencies and five duty-cycle values using the same LED visual stimulus. Figs. 7–10 show the box plots for the obtained FFT amplitudes for different duty-cycles and frequencies 7, 8, 9 and 10 Hz. Each box plot data consists of combined 1500 values from ten subjects (rather than individual subjects due to space constraints) with same duty-cycle value and with same visual stimulus frequency. The central line in the box shows the median value while the edges of the box are at 25 and 75th percentiles with the whisker values not being displayed. The Kruskal–Wallis tests, $\chi^2(\mathrm{d}f = 4;\ N = 1500)$ shows significant differences between different duty-cycle values for all ten participants: 7 Hz: $\chi^2 = 4.6 \times 10^{03}$; 8 Hz: $\chi^2 = 4.2 \times 10^{03}$; 9 Hz: $\chi^2 = 5.1 \times 10^{03}$; and 0 Hz: $\chi^2 = 5.3 \times 10^{03}$ with all the significance p-values very close to zero.

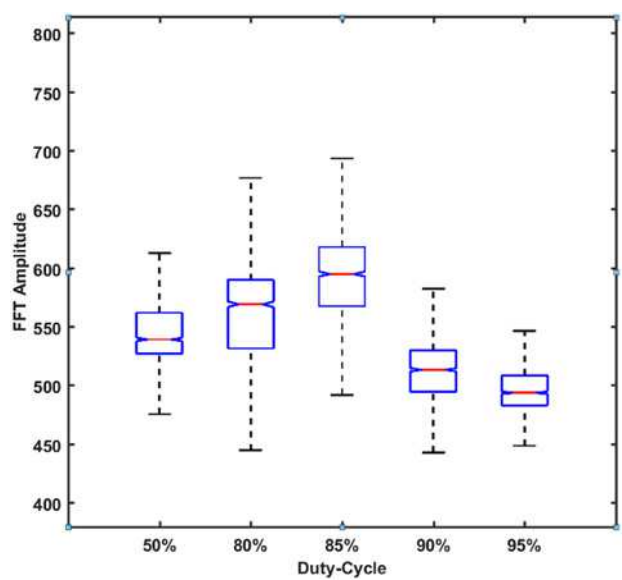

**Fig. 8** *Maximal FFT amplitude values from ten participants at 8 Hz*

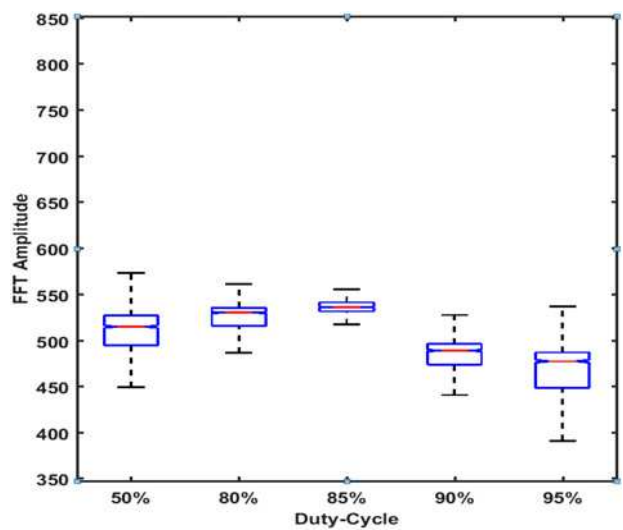

**Fig. 9** *Maximal FFT amplitude values from ten participants at 9 Hz*

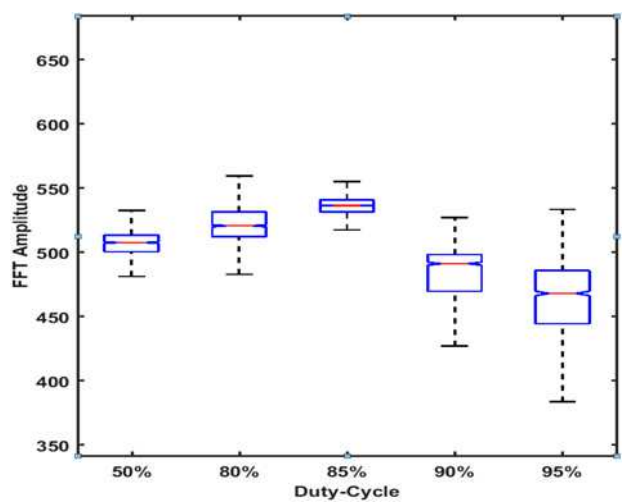

**Fig. 10** *Maximal FFT amplitude values from ten participants at 10 Hz*

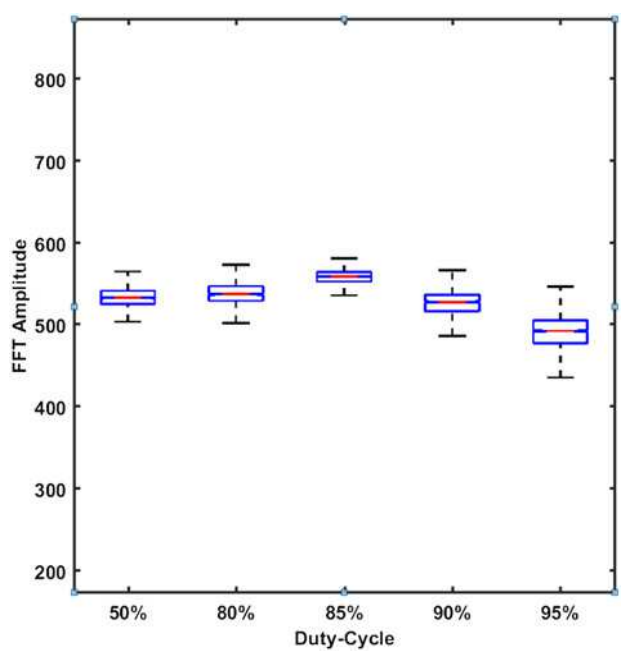

**Fig. 7** *Maximal FFT amplitude values from ten participants at 7 Hz*

**Table 1** Kruskal–Wallis mean-rank, average of maximal FFT [and standard deviation (SD)] (bold indicates the mean-rank values for the highest SSVEP response for duty-cycle at 85%)

| Participant | frequency | Duty-cycle | | | | | | | | | |
|---|---|---|---|---|---|---|---|---|---|---|---|
| | | 50% | | 80% | | 85% | | 90% | | 95% | |
| | | Mean-rank | Average ± SD | Mean-rank | Average ± SD | Mean-rank | Average ± SD | Mean-rank | Average ± SD | Mean-rank | Average ± SD |
| S1 | 7 Hz | 387.1 | 537.6 ± 4.7 | 387.3 | 538.1 ± 3.5 | **649.2** | 562.8 ± 8.6 | 377.4 | 535.1 ± 5.1 | 76.5 | 417.4 ± 8.7 |
| | 8 Hz | 366.8 | 581.5 ± 8.5 | 594.2 | 591.9 ± 6.3 | **597.1** | 651.1 ± 0.8 | 236.3 | 531.6 ± 6.9 | 83.1 | 490.2 ± 9.4 |
| | 9 Hz | 466.5 | 531.6 ± 2.6 | 545.5 | 535.9 ± 9.4 | **558.6** | 536.5 ± 0.3 | 188.9 | 490.1 ± 8.4 | 117.8 | 484.4 ± 7.5 |
| | 10 Hz | 404.5 | 503.8 ± 7.1 | 474.4 | 507.1 ± 2.2 | **646.9** | 533.2 ± 2.2 | 218.9 | 482.5 ± 9.3 | 132.6 | 487.8 ± 5.6 |
| S2 | 7 Hz | 330.3 | 532.1 ± 3.6 | 554.1 | 557.7 ± 8.6 | **598.2** | 561.9 ± 7.6 | 310.1 | 536.2 ± 3.1 | 85.0 | 486.2 ± 3.9 |
| | 8 Hz | 416.5 | 581.4 ± 9.1 | 489.8 | 591.3 ± 8.8 | **659.9** | 632.1 ± 9.9 | 209.5 | 538.1 ± 7.2 | 101.6 | 519.2 ± 1.4 |
| | 9 Hz | 411.6 | 525.9 ± 8.1 | 577.2 | 535.6 ± 0.1 | **581.1** | 534.0 ± 5.9 | 193.1 | 494.3 ± 9.9 | 114.3 | 489.0 ± 8.6 |
| | 10 Hz | 514.2 | 535.8 ± 4.3 | 516.5 | 531.5 ± 7.5 | **534.4** | 538.9 ± 4.8 | 235.7 | 514.0 ± 8.6 | 76.5 | 443.8 ± 6.3 |
| S3 | 7 Hz | 405.6 | 528.3 ± 4.6 | 485.5 | 529.2 ± 9.9 | **665.2** | 558.7 ± 8.1 | 244.5 | 509.4 ± 9.4 | 76.7 | 443.5 ± 2.4 |
| | 8 Hz | 376.1 | 537.6 ± 2.4 | 533.6 | 576.2 ± 7.6 | **636.7** | 593.7 ± 8.7 | 155.1 | 489.9 ± 7.1 | 155.9 | 489.7 ± 6.5 |
| | 9 Hz | 431.7 | 527.4 ± 1.1 | 560.6 | 531.1 ± 5.1 | **574.3** | 536.1 ± 0.7 | 191.7 | 492.2 ± 0.2 | 119.0 | 488.8 ± 8.6 |
| | 10 Hz | 340.7 | 503.4 ± 7.4 | 492.1 | 514.9 ± 5.8 | **670.9** | 535.7 ± 5.3 | 196.1 | 493.2 ± 7.5 | 177.6 | 489.6 ± 1.4 |
| S4 | 7 Hz | 463.5 | 535.1 ± 4.4 | 475.7 | 533.6 ± 3.8 | **621.5** | 547.8 ± 7.9 | 193.2 | 496.1 ± 3.7 | 123.3 | 484.9 ± 8.1 |
| | 8 Hz | 446.4 | 536.3 ± 8.3 | 437.1 | 534.3 ± 1.7 | **672.9** | 574.7 ± 8.3 | 184.2 | 510.1 ± 5.2 | 136.7 | 505.7 ± 1.8 |
| | 9 Hz | 441.1 | 582.1 ± 0.5 | 555.1 | 535.1 ± 5.9 | **567.1** | 537.8 ± 7.1 | 196.7 | 491.6 ± 0.1 | 117.2 | 482.4 ± 1.2 |
| | 10 Hz | 371.2 | 507.7 ± 0.2 | 516.1 | 520.1 ± 7.9 | **665.6** | 536.4 ± 6.1 | 195.6 | 492.7 ± 7.3 | 123.8 | 487.8 ± 5.6 |
| S5 | 7 Hz | 377.10 | 528.9 ± 7.1 | 436.3 | 535.7 ± 2.1 | **656.1** | 554.8 ± 6.5 | 328.4 | 526.7 ± 11.1 | 79.2 | 487.9 ± 9.1 |
| | 8 Hz | 416.63 | 537.9 ± 4.9 | 416.7 | 538.4 ± 0.8 | **670.9** | 591.9 ± 8.9 | 293.8 | 525.1 ± 6.1 | 79.3 | 491.6 ± 0.4 |
| | 9 Hz | 433.79 | 526.7 ± 9.3 | 561.4 | 535.5 ± 8.3 | **572.2** | 536.4 ± 0.3 | 180.4 | 494.6 ± 8.7 | 129.6 | 490.9 ± 0.9 |
| | 10 Hz | 401.51 | 510.2 ± 4.2 | 530.6 | 520.0 ± 7.9 | **636.1** | 536.2 ± 4.7 | 155.8 | 442.2 ± 6.3 | 150.4 | 443.4 ± 6.5 |
| S6 | 7 Hz | 346.1 | 535.7 ± 2.1 | 455.6 | 536.3 ± 8.2 | **667.9** | 559.9 ± 6.3 | 329.9 | 534.1 ± 3.9 | 77.9 | 536.2 ± 9.9 |
| | 8 Hz | 484.1 | 526.7 ± 6.1 | 459.3 | 525.6 ± 7.8 | **561.8** | 535.1 ± 9.2 | 211.9 | 501.2 ± 7.5 | 160.3 | 495.1 ± 5.8 |
| | 9 Hz | 379.1 | 505.8 ± 9.4 | 505.9 | 510.9 ± 5.7 | **669.9** | 533.1 ± 5.2 | 240.4 | 482.4 ± 9.5 | 82.2 | 465.6 ± 8.2 |
| | 10 Hz | 411.3 | 514.1 ± 7.6 | 491.8 | 520.3 ± 5.9 | **667.1** | 536.1 ± 6.1 | 182.8 | 492.1 ± 8.5 | 124.3 | 487.9 ± 7.4 |
| S7 | 7 Hz | 379.9 | 535.6 ± 5.7 | 391.1 | 535.8 ± 7.1 | **650.8** | 561.3 ± 8.2 | 368.7 | 534.2 ± 9.2 | 86.9 | 504.1 ± 9.3 |
| | 8 Hz | 353.7 | 509.9 ± 7.5 | 519.3 | 525.6 ± 6.3 | **636.4** | 537.1 ± 0.9 | 495.6 | 498.8 ± 9.3 | 156.1 | 489.8 ± 1.3 |
| | 9 Hz | 355.7 | 494.2 ± 0.8 | 526.2 | 537.1 ± 6.1 | **674.5** | 580.2 ± 3.2 | 194.1 | 477.7 ± 0.4 | 126.9 | 472.1 ± 1.3 |
| | 10 Hz | 345.9 | 509.1 ± 3.6 | 507.1 | 520.1 ± 7.9 | **658.7** | 535.5 ± 7.9 | 282.6 | 504.1 ± 7.6 | 83.1 | 473.4 ± 2.6 |
| S8 | 7 Hz | 404.6 | 535.2 ± 1.2 | 435.5 | 539.4 ± 7.6 | **638.9** | 559.2 ± 6.2 | 303.3 | 526.9 ± 33.2 | 95.1 | 495.8 ± 9.5 |
| | 8 Hz | 248.9 | 561.2 ± 6.7 | 495.9 | 592.8 ± 6.2 | **645.1** | 616.8 ± 9.6 | 397.8 | 546.3 ± 7.2 | 89.7 | 524.6 ± 1.4 |
| | 9 Hz | 414.1 | 511.8 ± 9.1 | 490.4 | 516.3 ± 5.7 | **670.3** | 532.1 ± 4.9 | 175.5 | 452.3 ± 7.9 | 127.2 | 448.3 ± 7.9 |
| | 10 Hz | 368.8 | 508.0 ± 3.2 | 594.8 | 535.5 ± 7.9 | **601.5** | 535.7 ± 5.3 | 236.2 | 492.7 ± 7.3 | 76.1 | 463.1 ± 6.8 |
| S9 | 7 Hz | 369.1 | 530.4 ± 6.8 | 424.3 | 538.1 ± 4.8 | **648.8** | 558.2 ± 8.1 | 332.9 | 529.8 ± 4.3 | 102.3 | 508.4 ± 3.7 |
| | 8 Hz | 380.1 | 538.3 ± 8.8 | 531.4 | 579.4 ± 8.2 | **659.7** | 611.3 ± 9.5 | 175.7 | 452.7 ± 8.5 | 130.4 | 482.5 ± 7.5 |
| | 9 Hz | 411.3 | 488.4 ± 9.2 | 490.7 | 495.1 ± 0.7 | **674.4** | 532.0 ± 5.2 | 215.3 | 410.3 ± 8.2 | 84.7 | 396.1 ± 7.1 |
| | 10 Hz | 372.6 | 501.2 ± 6.9 | 592.7 | 534.6 ± 7.8 | **607.4** | 531.2 ± 6.2 | 182.4 | 454.3 ± 9.8 | 122.3 | 447.3 ± 6.6 |
| S10 | 7 Hz | 388.5 | 532.4 ± 9.8 | 426.1 | 534.8 ± 1.2 | **659.1** | 559.8 ± 7.8 | 316.2 | 525.1 ± 6.5 | 87.4 | 498.2 ± 1.2 |
| | 8 Hz | 421.8 | 582.5 ± 8.5 | 482.1 | 594.8 ± 7.7 | **667.3** | 651.1 ± 0.8 | 229.6 | 528.1 ± 5.1 | 76.6 | 489.2 ± 9.1 |
| | 9 Hz | 242.9 | 486.4 ± 1.5 | 531.3 | 524.3 ± 5.8 | **658.7** | 531.1 ± 5.6 | 368.1 | 500.8 ± 12.2 | 76.3 | 439.4 ± 7.7 |
| | 10 Hz | 418.1 | 484.7 ± 6.8 | 459.6 | 490.6 ± 1.4 | **671.7** | 534.6 ± 6.3 | 249.1 | 469.5 ± 6.9 | 78.9 | 430.1 ± 8.7 |

Table 1 shows the mean ranks for all the different duty-cycles for ten subjects with four different frequencies. The mean ranks were individually computed for each frequency with different duty-cycle to analyse the influence of duty-cycle in SSVEP response. It can be observed that for all the subjects, the 85% duty-cycle gave the highest performance as compared with other duty-cycles. The SSVEP responses decreased after 85% duty-cycle even though the participants reported an increase in comfortability. The lowest SSVEP response was from 95% duty-cycle in all frequency ranges.

This paper also explored the level of comfortability for the users with different duty-cycles in all frequency ranges. A scale of 1–10 (10 being the most comfortable) was used for comparing the stimulus comfort with reduced visual fatigue for different duty-cycles. The participant response is shown as a bar graph in Fig. 11. Even though 90 and 95% duty-cycle shows the highest comfortability value, the SSVEP responses were less as compared with the responses from 85% duty-cycle. It is also known that the visual fatigue level also reduces with the increase in duty-cycle for visual stimulus. With 85% duty-cycle, even though the visual fatigue level (measured here with subject's comfortability rating) was more compared with higher duty-cycle, the SSVEP evoked had the maximal response and hence 85% duty-cycle is recommended for the visual stimulus.

## 4 Discussion

This paper has investigated the influence of different duty-cycle values in visual stimulus in eliciting SSVEP in brain for reducing visual fatigue and improving the comfortability of the user while giving improved response. The results were compared for frequencies 7, 8, 9 and 10 Hz with duty-cycle values of 50, 80, 85, 90 and 95%. The results show that there is an influence of duty-cycle in SSVEP responses which also agrees to previous studies [28, 29]. The data analysed for ten participants showed that the stimulus with 85% exhibited the highest response with minimal visual fatigue. The participants commented that they were able to gaze attentively for longer periods without discomfort producing better SSVEP responses. The duty-cycles above 85% showed a decrease in the SSVEP response that could likely be due to the very small $T_{OFF}$ period which might appear to the human eye to be constant light rather than flicker.

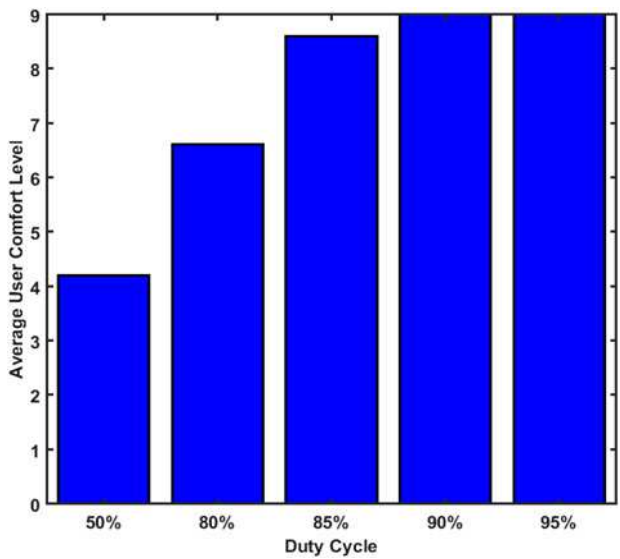

**Fig. 11** *Analysis of participant responses for visual stimulus comfortability*

All the participants felt there was improvement in comfortability to focus on visual stimulus when the duty-cycle increased as the stimulus was 'ON' for longer periods as compared with the conventional 50% ON/OFF duty-cycle which produces eye strain for prolonged usage. Furthermore, the use of various higher duty-cycles would also help in presenting multiple visual stimuli for classification purposes suited for BCI applications. Each stimulus could be customised with different duty-cycle values to override the influence of adjacent stimulus in multi-stimuli configuration using lower frequencies that are more influential for SSVEP. Further studies could explore the development of user adaptive duty-cycle where the stimulus hardware could automatically change the duty-cycle based on the real-time feedback from EEG response.